\documentclass[12pt]{article}

\newenvironment{wileykeywords}{\textsf{Keywords:}\hspace{\stretch{1}}}{\hspace{\stretch{1}}\rule{1ex}{1ex}}

\usepackage{amsmath,amssymb}
\usepackage{graphicx}% Include figure files
\usepackage{color}% Include colors for document elements
\usepackage{dcolumn}% Align table columns on decimal point
\usepackage{bm}% bold math
\usepackage[numbers,super,comma,sort&compress]{natbib}

\definecolor{background-color}{gray}{0.98}

\title{Computationally efficient double hybrid density functional theory using
dual basis methods}

\author{Jason N. Byrd\thanks{Quantum Theory Project, University of Florida, Gainesville, Florida 32611, United States}, John A. Montgomery, Jr.\thanks{Department of Physics, University of Connecticut, Storrs, CT 06269}}

\begin{document}

\maketitle

\begin{abstract}
We examine the application of the recently developed dual basis methods of Head-Gordon and co-workers to double hybrid density functional computations. Using the B2-PLYP, B2GP-PLYP, DSD-BLYP and DSD-PBEP86 density functionals, we assess the performance of dual basis methods for the calculation of conformational energy changes in C$_4$-C$_7$ alkanes and for the S22 set of noncovalent interaction energies. The dual basis methods, combined with resolution-of-the-identity second-order M{\o}ller-Plesset theory, are shown to give results in excellent agreement with conventional methods at a much reduced computational cost.
\end{abstract}

\begin{wileykeywords}
Double Hybrid Denstiy Functional Theory, Dual Basis SCF, S22, ACONF, DFT benchmark
%One, Two, Three, Four, Five.
%A list of five key words or phrases which best characterize the paper are required for indexing.
\end{wileykeywords}

\clearpage

%*****************Graphical Table of Contents******************** THIS IS MANDATORY *******************

\begin{figure}[h]
\centering
\colorbox{background-color}{
\fbox{
\begin{minipage}{1.0\textwidth}
\includegraphics[width=50mm,height=50mm]{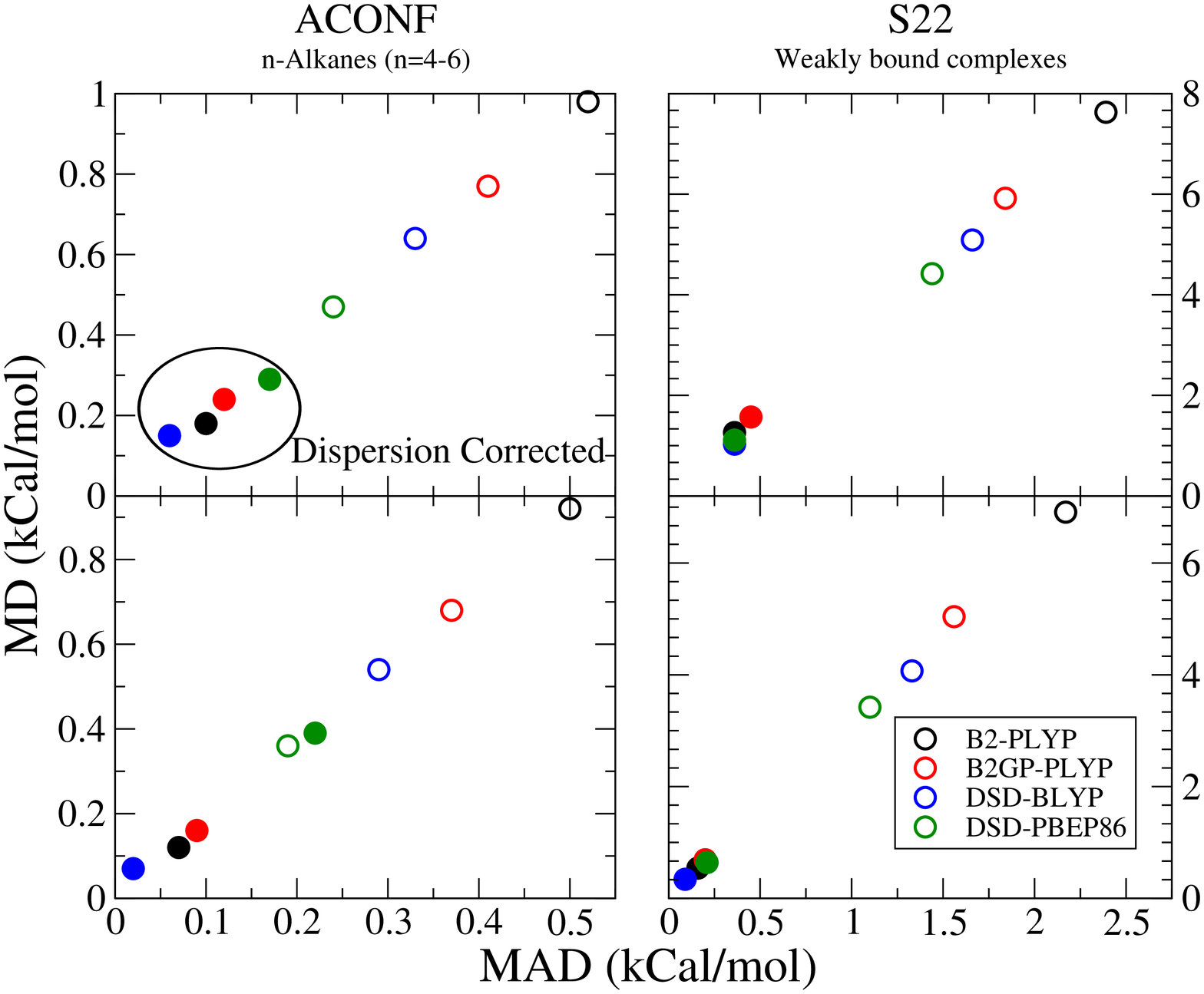} % Pick only one of the two styles by uncommenting the corresponding \includegraphics
\\
The maximum deviation (MD) plotted against mean average deviation for our dual basis double hybrid DFT benchmark test set results.
%(75 words.) Images for the graphical Table of Contents should capture the essence of a paper, displaying a figure, plot, or scheme that is central to the theme of the manuscript. The text of the graphical Table of Contents is meant for the non-specialist and should ideally contain no obscure jargon or mathematical symbols / equations, but should attempt to convey the gist of the paper in everyday terms, while remaining consistent with accepted standards of scientific literature.
\end{minipage}
}}
\end{figure}

% makes references listed with 1., 2., etc.  
  \makeatletter
  \renewcommand\@biblabel[1]{#1.}
  \makeatother

\bibliographystyle{apsrev}

\renewcommand{\baselinestretch}{1.5}
\normalsize

\clearpage

\section*{\sffamily \Large INTRODUCTION} % Not needed for rapid communications

Improving Kohn-Sham density functional theory and many body perturbation theory
by combining the two in some fashion
% by so called fifth rung methods\cite{perdew2005}
is an attractive idea garnering much recent attention\cite{gorling1993,gorling1994,bartlett2005,grimme2006a,verma2012c}.
Early difficulties in obtaining accurate results\cite{gorling1993} with Kohn-Sham (KS)
orbitals and second M{\o}ller-Plesset (MP2) perturbation theory (many body
perturbation theory with canonical self consistent field partitioning) were
addressed with the development of so called double hybrid (DH)
methods,\cite{grimme2006a,schwabe2008} where the idea is to use a mixture of
Hartree-Fock (HF) and density functional theory (DFT) exchange with a portion of
the correlation coming from DFT and the rest from an MP2 calculation using the
resulting Kohn-Sham orbitals.  This exchange-correlation partitioning is given by
\begin{equation}\label{dsddft}
E_{\rm XC} = (1-c_x)E_x^{\rm DFT} + c_xE_x^{\rm HF} + c_c E_c^{\rm DFT} +
c_{os}E_{os}^{\rm MP2} + c_{ss}E_{ss}^{\rm MP2} + E_D
\end{equation}
where $c_x$ is the amount of HF exchange, $c_c$ is the amount of DFT
correlation, $c_{os}$ and $c_{ss}$ are the mixing coefficients for the opposite
($\alpha\beta$) and same ($\alpha\alpha$) spin contributions to the MP2
correlation energy.  The -D2 (2006) and -D3 (2010) Grimme dispersion
correction\cite{grimme2006b,grimme2010} is included within the $E_D$ term as
desired.  For normal double hybrid methods such as B2-PLYP and B2GP-PLYP, the
same spin and opposite spin coefficients are constrained to be equal.  For DSD
methods such as DSD-BLYP\cite{kozuch2010} and DSD-PBEP86\cite{kozuch2011} this
constraint is lifted and the two components are optimized individually
in the same manner as spin component scaled MP2
(SCS-MP2).\cite{grimme2003,szabados2006,schwabe2008}  

The improvement of DH methods over standard DFT functionals has been well
documented\cite{goerigk2010,goerigk2011} for a wide variety of benchmark systems.
This general applicability and accuracy is expected from the DH methods, with
the added advantage of applicability to larger molecular systems where conventional
wavefunction methods such as coupled cluster theory\cite{bartlett2007} are
impractical.  Although second order M{\o}ller-Plesset perturbation theory is
perhaps the simplest useful method for computing the post-SCF electron correlation energy,
the formal $O(n^5)$ computational scaling becomes a very significant
bottleneck in large scale calculations.  The application of DH methods to larger
systems quickly reach this scaling bottleneck, resulting in the demand for
an alternative.  Similarly, the performance of the DFT portion of
the calculation demands an improvement due to scaling with system size.  The
purpose of this paper is to describe our proposed methodological changes to
circumvent the scaling limitations in standard DH methods.  Firstly, we address
the computational cost of standard DFT calculations by performing a dual basis self consistent
field (SCF) calculation.\cite{liang2004}  Following this the computational
cost of the MP2 portion of the calculation is reduced by using the
resolution-of-the-identity approximation\cite{feyereisen1993}.

The dual basis approach of Head-Gordon and co-workers\cite{liang2004,steele2006a,steele2007,steele2009}
is a method for reducing the computational cost of large basis SCF calculations. 
An SCF calculation in a small basis set
% is iterated to convergence, obtaining the small basis energy $E^{\rm small}_{\rm SCF}$
is iterated to convergence, providing the small basis energy $E_{\rm S}$
and molecular orbital (MO) coefficients $C_{S}$.
Approximate large basis MO coefficients $C_{L}$ are obtained by projection of
the small basis MO coefficients $C_{S}$ onto the large basis, found by
solving the linear equations
\begin{equation}
S_{L} C_{L} = S_{LS} C_{S}
% C_{L}^{\vphantom{-1}} = S_{L}^{-1} S_{LS}^{\vphantom{-1}} C_{S}^{\vphantom{-1}}
\end{equation}
where $S_{L}$ is the large basis overlap matrix and $S_{LS}$ is the rectangular
overlap matrix between the large and small basis.
Using the projected orbitals,
a new density matrix $P$ is formed, and a new Fock matrix, $F(P)$, is constructed
and diagonalized, providing a new density $P'.$
The density difference $\Delta P=P-P'$ is then used to compute a
correction to the small basis SCF energy from
\begin{equation}
\Delta E_{\rm L} = {\rm Tr} [\Delta PF(P)],
\end{equation}
and the dual basis SCF energy is obtained from
\begin{equation}
E_{\rm DB}=E_{\rm S} + \Delta E_{\rm L}.
\end{equation}

In this work we combine the dual basis SCF method with resolution-of-the-identity second-order M{\o}ller-Plesset theory.
The conventional expression for the MP2 correlation energy is
\begin{equation}
E_{\rm MP2} = -\sum_{ijab} \frac{(ia|jb)[2(ia|jb)-(ib|ja)]}{\epsilon_a+\epsilon_b-\epsilon_i-\epsilon_j}
\end{equation}
where $\epsilon_q$ are the SCF orbital eigenvalues and the indices $i,j\, (a,b)$
range over occupied (virtual) molecular orbitals. The occupied-virtual MOs are obtained from
a four index transformation of the AO basis electron repulsion integrals (ERIs)
\begin{equation}
(ia|jb) = 
\sum_\sigma C_{\sigma b} \sum_\lambda C_{\lambda j} 
\sum_\nu C_{\nu a} \sum_\mu C_{\mu i} 
(\mu\nu|\lambda\sigma)
\end{equation}
where greek indices range over AOs and $C_{\nu i}$ are MO coefficients. 
Resolution-of-the-identity methods seek to replace the ERIs
\begin{equation}
(\mu\nu|\lambda\sigma) = \int d{\bf r}_1  \int d{\bf r}_2 \; \phi_\mu ({\bf r}_1)
  \phi_\nu ({\bf r}_1) \frac{1}{r_{12}} \phi_\lambda ({\bf r}_2) \phi_\sigma ({\bf r}_2)
\end{equation}
by a sum of two and three center integrals
\begin{equation}
(\mu\nu|\lambda\sigma) \approx \sum_{lm} (\mu\nu|l)(l|m)^{-1}(m|\lambda\sigma)
\end{equation}
where $l,m$ are the indices of auxillary basis functions.
In practice, the size of the auxillary basis is typically 3-4 times larger
than the primary basis, leading to large speedups in the ERI calculation
and subsequent integral transformation.
% approach of Katouda {\it et al.}\cite{katouda2009} as is implemented in the
% GAMESS\cite{gamess1993,gamess2005} program. 
% Here the MP2 correlation energy is
% \begin{equation}
% E^{\rm large}_{\rm MP2} = 
% -\sum_{ijab} \frac{(ia|jb)[2(iaj|b)-(ia|bj)]}{\epsilon_a+\epsilon_b-\epsilon_i-\epsilon_j}
% \end{equation}
% where $\epsilon_q$ are the orbital eigenvalues, the indicies $ij (ab)$
% range over occupied (virtual) molecular orbitals and $(ia|jb)$ are spatial two
% electron integrals.  The rate limiting step in any
% MP2 calculation is the four index transformation from the AO (atomic orbital) basis to the MO (molecular
% orbital) basis.
% which scales formally as $ON^4$, where $O$ is the number of occupied orbitals and $N$ is the number of AOs.
% \begin{equation}
% (ia|jb) = 
% \sum_\sigma C_{\sigma b} \sum_\lambda C_{\lambda j} 
% \sum_\nu C_{\nu a} \sum_\mu C_{\mu i} 
% (\mu\nu|\lambda\sigma)
% \end{equation}
% where greek indices range over AOs and $C_{\nu i}$ are MO coefficients. 
% The RI scheme approximates the MO basis integrals using an
% auxilary basis set by\cite{katouda2009}
% \begin{equation}
% (ij|ab) = \sum_n B^{ia}_n B^{jb}_n,
% \end{equation}
% with
% \begin{equation}
% B^{ia}_n = \sum_\ell L^{-1}_{n\ell} \sum_\nu C_{\nu a} \sum_\mu (\mu\nu|\ell)C_{\mu i}. 
%\end{equation}
% Here $(\mu\nu|\ell)$ are three center integrals to be evaluated and $L^{-1}_{n\ell}$ is
% the Cholesky decomposition of $(\ell|m)$. 
% We defer further description of the
% algorithm and computational implementation of the GAMESS RI-MP2 to the original
% authors.\cite{katouda2009}

We note that MP2 calculations using dual basis SCF
orbitals formally have a singles contribution as the large basis density
is not completely relaxed and Brillouin's theorem therefore does not hold.
However Steele {\it et al.}\cite{steele2006a} have shown that the contribution of
the singles term is very small, therefore it is neglected in this work.
% Since the construction of the
% matrix $F'$ from the new density $P'$, this singles term is then given by
% \begin{equation}
% E^{\rm large}_{\rm singles} = \sum_{ai}
% \frac{F_{ai}^{'2}}{\epsilon_a-\epsilon_i}.
% \end{equation}
% singles correction involves a new Fock matrix construction the total SCF
% computational time is nearly doubled for what we find to be a small numerical
% correction.\cite{steele2006a}  As such this term is neglected in our
% implementation.

\section*{\sffamily \Large METHODOLOGY}

The large basis set used for dual basis calculations in this work are the
standard correlation consistent
Dunning\cite{dunning1989,kendall1992,woon1993,woon1994,woon1995,wilson1996}
triple zeta basis sets with (aug-cc-pVTZ) and without (cc-pVTZ) diffuse
functions.  The corresponding small basis sets are pruned sets from these the
larger (aug-) cc-pVTZ basis sets using the prescription given by Steele {\it et
al.}\cite{steele2006a,steele2009} which we denote r-cc-pVTZ and r-aug-cc-pVTZ in
this paper.  For r-cc-pVTZ only the $f$ functions are dropped, except in the
case of hydrogen and helium where in addition to the $s$ functions the tightest
$p$ function is kept.  The pruning scheme is similar for the r-aug-cc-pVTZ set
where in addition to the $f$ functions being dropped, the diffuse $d$ is also
dropped.  In the hydrogen and helium set all $s$ functions are kept as well as
the two tightest $p$ functions, all others are dropped.  In all RIMP2
calculations, we used the fitting basis sets from Weigend {\it et
al.}\cite{weigend2002}
% Dual basis pairings exist also for the quadruple zeta basis 
% sets (cc-pVQZ etc.), however the integral program in the GAMESS package is
% limited to including angular momentum through $g$ functions which necessarily
% excludes the use of the associated quadruple zeta fitting basis sets which
% contain $h$ functions.
For all DFT calculations in this work we make use of
the so called JANS=2 grid in GAMESS, which is a 155 radial Euler-MacLaurin
quadrature and pruned 974 point Lebedev grid.  Tests with a larger grid did not
lead to any appreciable differences in either conformer or dissociation
energies.  

To assess the usefullness of using dual basis SCF in a DH method, we
make use of the S22 set of non-covalently bonded molecules\cite{jurecka2006} and the ACONF set of
alkane (C$_4$-C$_6$) conformers\cite{gruzman2009}.  Reference geometries and
{\it ab initio}  dissociation energies for the S22 are taken from Jure\v{c}ka
{\it et al.}\cite{jurecka2006} and Takatani {\it et al.}\cite{takatani2010}
respectively.  For the ACONF set both reference geometries and {\it ab initio} conformer
energies are taken from Gruzman {\it et al.}\cite{gruzman2009}  Dissociation and
conformer energies were computed within the frozen core approximation
(core-valence energy is implicit within the DFT framework, but is explicitly
excluded from the MP2 correlation energy) as is usual for these sets.
All electronic structure calculations in this work were done using 
a locally modified version of the GAMESS
quantum chemistry program package\cite{gamess1993,gamess2005} in which we have
implemented the dual basis SCF scheme outlined in the introdction.
Throughout this paper we use RMS to mean "root mean square," MAD to mean "mean
average deviation," and MD to mean "maximum deviation." 

\section*{\sffamily \Large RESULTS}
\subsection*{\sffamily \large ALKANE CONFORMER ORDER AND RELATIVE ENERGIES}

The DB-RIDH results for the ACONF set of C$_4$-C$_6$ alkanes are presented in
Tables \ref{aconfcct} and \ref{aconfacct}, with and without the Grimme dispersion
correction.  Our reference {\it ab initio} data for the ACONF set comes from the
W1h-val results of Gruzman {\it et al.}\cite{gruzman2009} The W1h-val method
incorporates SCF and CCSD(T) (coupled cluster with perturbative singles and
doubles with perturbative triples\cite{Pople87}) methods where the final SCF and CCSD
energies are obtained using W2 extrapolation from the cc-pV{T,Q}Z basis sets
and the (T) contribution extrapolated from the cc-pV{D,T}Z basis set.  In
addition to the now standard ACONF set, we have also computed the conformer
energies for n-heptane; the results can be found in Tables \ref{heptcct} and
\ref{heptacct}.\footnote{Results for even longer alkane chains can be found in Byrd {\it
et al.}\cite{byrd2014-a}}  The {\it ab initio} n-heptane results from Gruzman {\it et
al.}\cite{gruzman2009} are at the slightly lower MP2:CC level of theory
(MP2/cc-pVQZ + [CCSD(T)/cc-pVDZ - MP2/cc-pVDZ] at a MP2/cc-pVDZ reference
geometry).  For n-butane through n-hexane (the ACONF set) the differences
between W1h-val and MP2:CC are reported\cite{gruzman2009} to be on the order of
0.01 kcal/mol, which is more than sufficiently accurate for our purposes.  We find
the RMS and MAD relative to the {\it ab initio}\cite{gruzman2009} values
to be around $0.5$ kcal/mol for all functionals used, with almost a
factor of two improvement for the DSD methods over the standard B2-PLYP.
Incuding Grimme dispersion further improves the results to obtain nearly
indistinguishable values compared to {\it ab initio}.  As expected for alkane
conformers, the differences between the cc-pVTZ and the diffuse aug-cc-pVTZ
basis sets are negligible.  

The energetic ordering of each conformer is as important as the relative
eergies.  For both basis sets the DH methods predict the same conformer ACONF
ordering excluding the G$^+$T$^+$G$^-$-GGG conformers\footnote{It should be
noted that the energy difference between the G$^+$T$^+$G$^-$ and GGG conformers
is $0.05$ kcal/mol, the smallest of the test set.}.  Only the DSD methods
including Grimme dispersion agree with the {\it ab initio} ordering when using
the cc-pVTZ basis set.  Including diffuse functions with the aug-cc-pVTZ basis
set brings every DH method into agreement with {\it ab initio} so long as Grimme
dispersion is included.  This same conformer ordering error is seen in standard DH
calculations\cite{goerigk2011} and so is not an indication of a problem with
either using dual basis or RIMP2.  For the case of n-heptane, the DB-RIDH
calculations with and without Grimme dispersion have small number of non-systematic 
conformer ordering discrepancies.  As the magnetude of the error in these
discrepancies (errors near 0.01 kcal/mol) are the same size as the
reported errors in MP2:CC we do not attach any particular significance to the
differences in conformer ordering.

\subsection*{\sffamily \large S22 NON-COVALENT INTERACTION ENERGIES}

Tables \ref{s22cct} and \ref{s22acct} show the DB-RIHD results (with and without Grimme
dispersion) for the S22 set of non-covalently bonded molecules.  In the
absence of a dispersion correction, the RMS and MAD values for both the cc-pVTZ
and aug-cc-pVTZ basis sets relative to the benchmark CBS($\Delta$a(DT)Z) {\it ab
initio} results\cite{takatani2010} range from 1 to 3 kcal/mol with the DSD
methods performing the best overall.  Inclusion of Grimme dispersion brings the
agreement for all DH methods to better than half a kcal/mol for the
cc-pVTZ basis set.  As expected considering the number of polar molecules
included within the S22 set, the inclusion of diffuse functions in the
aug-cc-pVTZ basis reduces the RMS and MAD error even more.  

All DH methods and further dispersion corrections are parameterized methods
that are optimized using various test sets often employing QZ level basis
sets\cite{grimme2006a,karton2008} or TZ level basis sets other than the Dunning
type correlation sets (such as the PC-n\cite{jensen2001,jensen2002} and
def2\cite{weigend2003,weigend2005} basis sets).  As such, care must be made when
using basis sets other than the ones used for parametrization.  The DH benchmark
S22 calculations of Goerigk {\it et al.}\cite{goerigk2011} give MAD and RMS
values of $0.27$ and $0.33$ kcal/mol for B2-PLYP-D3/def2-QZVPP and $0.28$
and $0.34$ kcal/mol for DSD-BLYP-D3/def2-QZVPP.  Our own MAD and RMS
values of $0.36$ and $0.48$ kcal/mol for B2-PLYP-D3/c-pVTZ and $0.36$ and
$0.46$ kcal/mol for DSD-BLYP-D3/cc-pVTZ are then entirely acceptable.  It
should be noted that our results of $0.16$ and $0.21$ kcal/mol for
B2-PLYP-D3/aug-cc-pVTZ and $0.09$ and $0.13$ kcal/mol for
DSD-BLYP-D3/aug-cc-pVTZ are a marked improvement over not only the cc-pVTZ
results but also the benchmark def2-QZVPP Goerigk {\it et al.}\cite{goerigk2011}
values, illustrating the need for diffuse functions to accurately describe 
non-covalent bonds involving polar molecules.

%The errors seen here
%are completely consistent with the standard DH errors reported by Goerigk {\it
%et al.}\cite{goerigk2011} which leads us to conclude that the accuracy penalty
%of using dual basis (and RI) for DH methods is negligible in the case of
%dissociation energies.  

%Addressing just this thought is the DuT method from Chan {\it et
%al.}\cite{chan2011}, a reoptimized DSD-BLYP method designed to
%use with the aug'-cc-pVTZ (' denotes no diffuse functions on hydrogen).
%To assertain how much error we accrue using the cc-pVTZ basis sets we have recalculated the S22 set
%using the DuT method, where we find no appreciable difference between the DuT
%and DSD-BLYP methods using the same aug' basis set.

\section*{\sffamily \Large CONCLUSIONS}

We have shown that the use of dual basis and resolution-of-the-identity methods can speed
up double hybrid DFT computations without significant loss in accuracy.
Results are presented for alkane conformational energy differences and
non-covalent interaction energies using 
the B2-PLYP, B2GP-PLYP, DSD-BLYP and DSD-PBEP86 double hybrid density functionals.

%((Place Conclusions here.))

\subsection*{\sffamily \large ACKNOWLEDGMENTS}

We are grateful to the Booth Engineering Center for Advanced Technology at
the University of Connecticut for the use of its computational facilities.

%((Place Acknowledgments here))

%((Additional Supporting Information may be found in the online version of this article.))

\clearpage

%%%%%%%%%%%%%%%%%%%%%%%%%%%%%%%%%%%%%%%%%%%%%%%%%%%%%%%%%%%%%%%%%%%%%%%%%%%%%%%%%
% BIBLIOGRAPHY

%\bibliography{l2}   % Produces the bibliography via BibTeX.

%%%%%%%%%%%%%%%%%%%%%%%%%%%%%%%%%%%%%%%%%%%%%%%%%%%%%%%%%%%%%%%%%%%%%%%%%%%%%%%%%

\clearpage

\begin{table}\footnotesize
\caption{\label{aconfcct}Dual basis resolution-of-the-identity double hybrid cc-pVTZ conformer energies for the
ACONF C$_4$-C$_6$ test set in kcal/mol.  {\it Ab initio} reference
energies are the W1h-val values from Gruzman {\it et al.}\cite{gruzman2009}
}
\begin{tabular}{lrrrrrrrrr}
\hline
\hline
\multicolumn{1}{c}{cc-pVTZ}           
& \multicolumn{2}{c}{B2-PLYP} & \multicolumn{2}{c}{B2GP-PLYP} & \multicolumn{2}{c}{DSD-BLYP} & \multicolumn{2}{c}{DSD-PBEP86} 
& \multicolumn{1}{c}{\it ab initio} \\
& & \multicolumn{1}{c}{$-D3$} & & \multicolumn{1}{c}{$-D3$} & & \multicolumn{1}{c}{$-D3$} & & \multicolumn{1}{c}{$-D2$} \\
\hline
\multicolumn{10}{c}{n-butane}\\
T & 0.00 & 0.00 & 0.00 & 0.00 & 0.00 & 0.00 & 0.00 & 0.00 & 0.00 \\
G & 0.79 & 0.66 & 0.74 & 0.66 & 0.72 & 0.64 & 0.68 & 0.54 & 0.60 \\
\multicolumn{10}{c}{n-pentane}\\
TT & 0.00 & 0.00 & 0.00 & 0.00 & 0.00 & 0.00 & 0.00 & 0.00 & 0.00 \\
TG & 0.81 & 0.66 & 0.76 & 0.66 & 0.73 & 0.64 & 0.69 & 0.53 & 0.61 \\
GG & 1.51 & 1.08 & 1.37 & 1.07 & 1.29 & 1.01 & 1.19 & 0.79 & 0.96 \\
GX$^-$ & 3.30 & 2.91 & 3.22 & 2.96 & 3.15 & 2.91 & 3.05 & 2.68 & 2.81 \\
\multicolumn{10}{c}{n-hexane}\\
TTT & 0.00 & 0.00 & 0.00 & 0.00 & 0.00 & 0.00 & 0.00 & 0.00 & 0.00 \\
GTT & 0.80 & 0.64 & 0.75 & 0.64 & 0.71 & 0.61 & 0.68 & 0.51 & 0.60 \\
TGT & 0.81 & 0.65 & 0.76 & 0.65 & 0.73 & 0.62 & 0.69 & 0.51 & 0.60 \\
TGG & 1.54 & 1.04 & 1.38 & 1.04 & 1.28 & 0.97 & 1.18 & 0.74 & 0.93 \\
GTG & 1.61 & 1.27 & 1.50 & 1.27 & 1.43 & 1.21 & 1.35 & 1.01 & 1.18 \\
G$^+$T$^+$G$^-$ & 1.67 & 1.39 & 1.58 & 1.40 & 1.53 & 1.35 & 1.46 & 1.13 & 1.30 \\
GGG & 2.23 & 1.43 & 1.96 & 1.42 & 1.82 & 1.31 & 1.66 & 0.96 & 1.25 \\
G$^+$X$^-$T$^+$ & 3.19 & 2.74 & 3.09 & 2.78 & 3.01 & 2.72 & 2.90 & 2.48 & 2.63 \\
T$^+$G$^+$X$^-$ & 3.26 & 2.83 & 3.16 & 2.87 & 3.09 & 2.81 & 2.98 & 2.57 & 2.74 \\
G$^+$X$^-$G$^-$ & 4.01 & 3.45 & 3.87 & 3.50 & 3.76 & 3.41 & 3.63 & 3.06 & 3.28 \\
X$^+$G$^-$G$^-$ & 3.99 & 3.19 & 3.80 & 3.26 & 3.67 & 3.16 & 3.51 & 2.80 & 3.08 \\
X$^+$G$^-$X$^+$ & 5.85 & 5.06 & 5.69 & 5.16 & 5.57 & 5.07 & 5.39 & 4.64 & 4.93 \\
\hline
RMS  & 0.59 & 0.11 & 0.46 & 0.14 & 0.37 & 0.07 & 0.27 & 0.19 &  \\
MAD  & 0.52 & 0.10 & 0.41 & 0.12 & 0.33 & 0.06 & 0.24 & 0.17 &  \\
MD   & 0.98 & 0.18 & 0.77 & 0.24 & 0.64 & 0.15 & 0.47 & 0.29 &  \\
%standard dev & 0.47 & 0.07 & 0.31 & 0.05 & 0.26 & 0.04 & 0.21 & 0.06 & 0.17 & 0.04 & 0.13 & 0.07 &  \\
\hline
\hline
\end{tabular}\end{table}

\begin{table}\footnotesize
\caption{\label{aconfacct}Dual basis resolution-of-the-identity double hybrid
aug-cc-pVTZ conformer energies for the
ACONF C$_4$-C$_6$ test set in kcal/mol.  {\it Ab initio} reference
energies are the W1h-val values from Gruzman {\it et al.}\cite{gruzman2009}
}
\begin{tabular}{lrrrrrrrrr}
\hline
\hline
\multicolumn{1}{c}{aug-cc-pVTZ}           
& \multicolumn{2}{c}{B2-PLYP} & \multicolumn{2}{c}{B2GP-PLYP} & \multicolumn{2}{c}{DSD-BLYP} & \multicolumn{2}{c}{DSD-PBEP86}
& \multicolumn{1}{c}{\it ab initio} \\
%& \multicolumn{1}{c}{W1h-val} \\
& & \multicolumn{1}{c}{$-D3$} & 
  & \multicolumn{1}{c}{$-D3$} & & \multicolumn{1}{c}{$-D3$} & & \multicolumn{1}{c}{$-D2$} \\
\hline
\multicolumn{10}{c}{n-butane}\\
T & 0.00 & 0.00 & 0.00 & 0.00 & 0.00 & 0.00 & 0.00 & 0.00 & 0.00 \\
G & 0.79 & 0.66 & 0.74 & 0.65 & 0.71 & 0.63 & 0.68 & 0.53 & 0.60 \\
\multicolumn{10}{c}{n-pentane}\\
TT & 0.00 & 0.00 & 0.00 & 0.00 & 0.00 & 0.00 & 0.00 & 0.00 & 0.00 \\
TG & 0.80 & 0.66 & 0.75 & 0.65 & 0.72 & 0.63 & 0.68 & 0.52 & 0.61 \\
GG & 1.49 & 1.05 & 1.34 & 1.04 & 1.25 & 0.97 & 1.15 & 0.75 & 0.96 \\
GX$^-$ & 3.27 & 2.88 & 3.18 & 2.92 & 3.11 & 2.86 & 3.00 & 2.63 & 2.81 \\
\multicolumn{10}{c}{n-hexane}\\
TTT & 0.00 & 0.00 & 0.00 & 0.00 & 0.00 & 0.00 & 0.00 & 0.00 & 0.00 \\
GTT & 0.79 & 0.63 & 0.74 & 0.63 & 0.70 & 0.60 & 0.66 & 0.49 & 0.60 \\
TGT & 0.81 & 0.65 & 0.75 & 0.64 & 0.72 & 0.61 & 0.68 & 0.50 & 0.60 \\
TGG & 1.51 & 1.01 & 1.34 & 1.00 & 1.24 & 0.93 & 1.13 & 0.69 & 0.93 \\
GTG & 1.59 & 1.25 & 1.48 & 1.25 & 1.40 & 1.18 & 1.32 & 0.97 & 1.18 \\
G$^+$T$^+$G$^-$ & 1.66 & 1.38 & 1.56 & 1.38 & 1.50 & 1.33 & 1.43 & 1.10 & 1.30 \\
GGG & 2.17 & 1.37 & 1.90 & 1.36 & 1.74 & 1.23 & 1.58 & 0.88 & 1.25 \\
G$^+$X$^-$T$^+$ & 3.15 & 2.69 & 3.04 & 2.73 & 2.95 & 2.67 & 2.84 & 2.42 & 2.63 \\
T$^+$G$^+$X- & 3.23 & 2.79 & 3.12 & 2.83 & 3.04 & 2.77 & 2.93 & 2.52 & 2.74 \\
G$^+$X$^-$G$^-$ & 3.97 & 3.40 & 3.82 & 3.44 & 3.70 & 3.35 & 3.55 & 2.99 & 3.28 \\
X$^+$G$^-$G$^-$ & 3.94 & 3.13 & 3.73 & 3.19 & 3.59 & 3.08 & 3.42 & 2.71 & 3.08 \\
X$^+$G$^-$X$^+$ & 5.78 & 4.99 & 5.61 & 5.08 & 5.47 & 4.98 & 5.29 & 4.53 & 4.93 \\
\hline
RMS  & 0.55 & 0.07 & 0.42 & 0.09 & 0.32 & 0.03 & 0.21 & 0.24 &  \\
MAD  & 0.50 & 0.07 & 0.37 & 0.09 & 0.29 & 0.02 & 0.19 & 0.22 &  \\
MD   & 0.92 & 0.12 & 0.68 & 0.16 & 0.54 & 0.07 & 0.36 & 0.39 &  \\
%standard dev & 0.45 & 0.07 & 0.30 & 0.04 & 0.24 & 0.03 & 0.18 & 0.04 & 0.14 & 0.02 & 0.10 & 0.10 &  \\
\hline
\hline
\end{tabular}\end{table}

\begin{table}\scriptsize
\caption{\label{heptcct}Dual basis resolution-of-the-identity double hybrid cc-pVTZ conformer energies for the
n-heptane conformers in kcal/mol.  {\it Ab initio} reference
energies are the MP2:CC values from Gruzman {\it et al.}\cite{gruzman2009}
}
\begin{tabular}{lrrrrrrrrr}
\hline\hline
\multicolumn{1}{c}{cc-pVTZ}           
& \multicolumn{2}{c}{B2-PLYP} & \multicolumn{2}{c}{B2GP-PLYP} & \multicolumn{2}{c}{DSD-BLYP} & \multicolumn{2}{c}{DSD-PBEP86}
& \multicolumn{1}{c}{\it ab initio}\\
& & \multicolumn{1}{c}{$-D3$} & & \multicolumn{1}{c}{$-D3$} & & \multicolumn{1}{c}{$-D3$} & & \multicolumn{1}{c}{$-D2$} \\
\hline
\multicolumn{10}{c}{n-heptane}\\
TTTT & 0.00 & 0.00 & 0.00 & 0.00 & 0.00 & 0.00 & 0.00 & 0.00 & 0.00 \\
TTTG$^-$ & 0.78 & 0.62 & 0.73 & 0.63 & 0.70 & 0.60 & 0.67 & 0.51 & 0.59 \\
TTG$^-$T & 0.79 & 0.61 & 0.73 & 0.62 & 0.70 & 0.59 & 0.66 & 0.49 & 0.57 \\
TTG$^-$G$^-$ & 1.44 & 0.97 & 1.29 & 0.97 & 1.21 & 0.91 & 1.12 & 0.71 & 0.92 \\
TG$^+$G$^+$T & 1.47 & 0.96 & 1.31 & 0.97 & 1.22 & 0.90 & 1.13 & 0.70 & 0.90 \\
TG$^+$TG$^+$ & 1.58 & 1.23 & 1.47 & 1.23 & 1.40 & 1.18 & 1.33 & 0.99 & 1.14 \\
G$^+$TTG$^+$ & 1.55 & 1.23 & 1.45 & 1.23 & 1.39 & 1.19 & 1.32 & 1.01 & 1.16 \\
G$^+$TTG$^-$ & 1.56 & 1.24 & 1.47 & 1.25 & 1.40 & 1.20 & 1.34 & 1.02 & 1.17 \\
TG$^+$TG$^-$ & 1.65 & 1.36 & 1.57 & 1.38 & 1.51 & 1.33 & 1.45 & 1.13 & 1.29 \\
TG$^+$G$^+$G$^+$ & 2.11 & 1.31 & 1.86 & 1.32 & 1.72 & 1.22 & 1.58 & 0.92 & 1.22 \\
G$^+$TG$^+$G$^+$ & 2.22 & 1.53 & 2.01 & 1.54 & 1.88 & 1.45 & 1.75 & 1.16 & 1.44 \\
G$^+$TG$^-$G$^-$ & 2.30 & 1.71 & 2.12 & 1.73 & 2.02 & 1.64 & 1.90 & 1.35 & 1.64 \\
G$^+$G$^+$G$^+$G$^+$ & 2.76 & 1.70 & 2.43 & 1.70 & 2.25 & 1.57 & 2.05 & 1.17 & 1.55 \\
TTX$^-$G$^+$ & 3.15 & 2.69 & 3.05 & 2.75 & 2.98 & 2.69 & 2.87 & 2.48 & 2.63 \\
TTG$^-$X$^+$ & 3.20 & 2.75 & 3.10 & 2.81 & 3.03 & 2.75 & 2.93 & 2.54 & 2.71 \\
TG$^+$X$^-$T & 3.10 & 2.62 & 2.99 & 2.67 & 2.91 & 2.61 & 2.80 & 2.39 & 2.56 \\
TG$^+$G$^+$X$^-$ & 3.88 & 3.02 & 3.68 & 3.10 & 3.55 & 3.00 & 3.38 & 2.67 & 2.96 \\
TX$^+$G$^-$G$^-$ & 3.78 & 2.97 & 3.59 & 3.05 & 3.46 & 2.96 & 3.31 & 2.63 & 2.92 \\
G$^+$TX$^+$G$^-$ & 3.97 & 3.32 & 3.82 & 3.39 & 3.71 & 3.31 & 3.56 & 2.99 & 3.21 \\
G$^+$TG$^+$X$^-$ & 4.00 & 3.40 & 3.86 & 3.46 & 3.75 & 3.38 & 3.62 & 3.08 & 3.30 \\
G$^+$TX$^-$G$^+$ & 4.03 & 3.41 & 3.88 & 3.46 & 3.77 & 3.38 & 3.62 & 3.07 & 3.29 \\
TG$^+$X$^+$G$^-$ & 3.94 & 3.35 & 3.80 & 3.42 & 3.70 & 3.34 & 3.57 & 3.03 & 3.26 \\
TG$^+$X$^-$G$^-$ & 3.90 & 3.32 & 3.77 & 3.38 & 3.66 & 3.30 & 3.53 & 2.99 & 3.19 \\
G$^+$G$^+$G$^+$X$^-$ & 4.62 & 3.56 & 4.36 & 3.64 & 4.19 & 3.52 & 3.99 & 3.11 & 3.54 \\
G$^+$G$^+$X$^-$G$^-$ & 4.57 & 3.55 & 4.32 & 3.64 & 4.16 & 3.52 & 3.97 & 3.09 & 3.45 \\
TX$^+$G$^-$X$^+$ & 5.59 & 4.69 & 5.41 & 4.81 & 5.28 & 4.71 & 5.09 & 4.33 & 4.61 \\
G$^+$X$^-$X$^-$G$^+$ & 5.92 & 4.98 & 5.70 & 5.06 & 5.54 & 4.95 & 5.32 & 4.53 & 4.78 \\
G$^+$X$^+$G$^-$X$^+$ & 6.50 & 5.56 & 6.31 & 5.69 & 6.17 & 5.59 & 5.97 & 5.12 & 5.42 \\
L$^+$G$^-$X$^-$G$^+$ & 7.02 & 6.17 & 6.85 & 6.30 & 6.71 & 6.19 & 6.49 & 5.74 & 5.97 \\
X$^+$G$^-$G$^-$X$^+$ & 7.10 & 6.22 & 6.98 & 6.40 & 6.86 & 6.32 & 6.68 & 5.92 & 6.28 \\
\hline
RMS                  & 0.76 & 0.10 & 0.59 & 0.15 & 0.48 & 0.08 & 0.34 & 0.24 & \\
MAD                  & 0.69 & 0.08 & 0.54 & 0.13 & 0.44 & 0.06 & 0.31 & 0.23 & \\
MD                   & 1.21 & 0.20 & 0.92 & 0.33 & 0.76 & 0.22 & 0.55 & 0.43 & \\
\hline
\hline
\end{tabular}\end{table}

\begin{table}\scriptsize
\caption{\label{heptacct}Dual basis resolution-of-the-identity double hybrid
aug-cc-pVTZ conformer energies for the
n-heptane conformers in kcal/mol.  {\it Ab initio} reference
energies are the MP2:CC values from Gruzman {\it et al.}\cite{gruzman2009}
}
\begin{tabular}{lrrrrrrrrr}
\hline
\hline
\multicolumn{1}{c}{cc-pVTZ}           
& \multicolumn{2}{c}{B2-PLYP} & \multicolumn{2}{c}{B2GP-PLYP} & \multicolumn{2}{c}{DSD-BLYP} & \multicolumn{2}{c}{DSD-PBEP86} 
& \multicolumn{1}{c}{\it ab initio}\\
& & \multicolumn{1}{c}{$-D3$} & & \multicolumn{1}{c}{$-D3$} & & \multicolumn{1}{c}{$-D3$} & & \multicolumn{1}{c}{$-D2$} \\
\hline
\multicolumn{9}{c}{n-heptane}\\
TTTT & 0.00 & 0.00 & 0.00 & 0.00 & 0.00 & 0.00 & 0.00 & 0.00 & 0.00\\
TTTG$^-$ & 0.77 & 0.62 & 0.72 & 0.62 & 0.69 & 0.59 & 0.66 & 0.50 & 0.59\\
TTG$^-$T & 0.78 & 0.61 & 0.73 & 0.61 & 0.69 & 0.58 & 0.65 & 0.48 & 0.57\\
TTG$^-$G$^-$ & 1.42 & 0.94 & 1.26 & 0.94 & 1.17 & 0.87 & 1.08 & 0.67 & 0.92\\
TG$^+$G$^+$T & 1.44 & 0.93 & 1.28 & 0.93 & 1.18 & 0.85 & 1.08 & 0.65 & 0.90\\
TG$^+$TG$^+$ & 1.56 & 1.22 & 1.45 & 1.21 & 1.38 & 1.15 & 1.30 & 0.96 & 1.14\\
G$^+$TTG$^+$ & 1.54 & 1.22 & 1.44 & 1.22 & 1.37 & 1.16 & 1.30 & 0.98 & 1.16\\
G$^+$TTG$^-$ & 1.55 & 1.22 & 1.45 & 1.23 & 1.38 & 1.18 & 1.31 & 0.99 & 1.17\\
TG$^+$TG$^-$ & 1.64 & 1.35 & 1.55 & 1.36 & 1.49 & 1.31 & 1.42 & 1.11 & 1.29\\
TG$^+$G$^+$G$^+$ & 2.06 & 1.26 & 1.80 & 1.26 & 1.65 & 1.15 & 1.50 & 0.84 & 1.22\\
G$^+$TG$^+$G$^+$ & 2.18 & 1.49 & 1.96 & 1.50 & 1.82 & 1.39 & 1.68 & 1.09 & 1.44\\
G$^+$TG$^-$G$^-$ & 2.27 & 1.68 & 2.08 & 1.69 & 1.97 & 1.60 & 1.85 & 1.30 & 1.64\\
G$^+$G$^+$G$^+$G$^+$ & 2.69 & 1.63 & 2.35 & 1.63 & 2.15 & 1.47 & 1.95 & 1.07 & 1.55\\
TTX$^-$G$^+$ & 3.11 & 2.65 & 3.00 & 2.70 & 2.92 & 2.63 & 2.81 & 2.42 & 2.63\\
TTG$^-$X$^+$ & 3.16 & 2.72 & 3.07 & 2.77 & 2.98 & 2.71 & 2.88 & 2.49 & 2.71\\
TG$^+$X$^-$T & 3.05 & 2.58 & 2.94 & 2.62 & 2.85 & 2.55 & 2.74 & 2.33 & 2.56\\
TG$^+$G$^+$X$^-$ & 3.82 & 2.96 & 3.61 & 3.02 & 3.45 & 2.91 & 3.28 & 2.57 & 2.96\\
TX$^+$G$^-$G$^-$ & 3.71 & 2.91 & 3.52 & 2.97 & 3.37 & 2.87 & 3.21 & 2.53 & 2.92\\
G$^+$TX$^+$G$^-$ & 3.91 & 3.26 & 3.75 & 3.33 & 3.63 & 3.23 & 3.47 & 2.90 & 3.21\\
G$^+$TG$^+$X$^-$ & 3.96 & 3.36 & 3.81 & 3.41 & 3.70 & 3.32 & 3.56 & 3.01 & 3.30\\
G$^+$TX$^-$G$^+$ & 3.98 & 3.36 & 3.82 & 3.40 & 3.70 & 3.31 & 3.55 & 3.00 & 3.29\\
TG$^+$X$^+$G$^-$ & 3.89 & 3.31 & 3.75 & 3.36 & 3.63 & 3.27 & 3.50 & 2.96 & 3.26\\
TG$^+$X$^-$G$^-$ & 3.86 & 3.28 & 3.72 & 3.33 & 3.60 & 3.24 & 3.46 & 2.92 & 3.19\\
G$^+$G$^+$G$^+$X$^-$ & 4.54 & 3.49 & 4.27 & 3.56 & 4.08 & 3.42 & 3.87 & 3.00 & 3.54\\
G$^+$G$^+$X$^-$G$^-$ & 4.49 & 3.48 & 4.23 & 3.55 & 4.05 & 3.41 & 3.85 & 2.97 & 3.45\\
TX$^+$G$^-$X$^+$ & 5.50 & 4.60 & 5.31 & 4.71 & 5.16 & 4.59 & 4.97 & 4.20 & 4.61\\
G$^+$X$^-$X$^-$G$^+$ & 5.82 & 4.89 & 5.59 & 4.96 & 5.41 & 4.82 & 5.18 & 4.40 & 4.78\\
G$^+$X$^+$G$^-$X$^+$ & 6.42 & 5.48 & 6.22 & 5.60 & 6.05 & 5.47 & 5.85 & 5.00 & 5.42\\
L$^+$G$^-$X$^-$G$^+$ & 6.93 & 6.07 & 6.74 & 6.18 & 6.57 & 6.05 & 6.35 & 5.59 & 5.97\\
X$^+$G$^-$G$^-$X$^+$ & 7.06 & 6.17 & 6.92 & 6.34 & 6.77 & 6.23 & 6.58 & 5.83 & 6.28\\
\hline
RMS & 0.71 & 0.06 & 0.53 & 0.09 & 0.41 & 0.04 & 0.26 & 0.32 & \\
MAD & 0.65 & 0.05 & 0.49 & 0.08 & 0.37 & 0.03 & 0.24 & 0.30 & \\
MD & 1.14 & 0.11 & 0.81 & 0.21 & 0.63 & 0.12 & 0.43 & 0.54 & \\
\hline
\hline
\end{tabular}\end{table}

\begin{table}\footnotesize
\caption{\label{s22cct}Dual basis resolution-of-the-identity double hybrid
aug-cc-pVTZ conformer energies for the
S22 test set of non-covalent bonded molecules in kcal/mol.  {\it Ab
initio} reference energies are the CBS($\Delta$a(DT)Z) values
from Takatani {\it et al.}\cite{takatani2010}
}
\begin{tabular}{lrrrrrrrrr}
\hline
\hline
\multicolumn{1}{c}{cc-pVTZ}           
& \multicolumn{2}{c}{B2-PLYP} & \multicolumn{2}{c}{B2GP-PLYP} & \multicolumn{2}{c}{DSD-BLYP} & \multicolumn{2}{c}{DSD-PBEP86} 
& \multicolumn{1}{c}{\it ab initio} \\
& & \multicolumn{1}{c}{$-D3$} & 
  & \multicolumn{1}{c}{$-D3$} & & \multicolumn{1}{c}{$-D3$} & & \multicolumn{1}{c}{$-D2$} \\
\hline
\multicolumn{10}{c}{H-bonded complexes}\\
$\rm(NH_3)_2$      & -2.58 & -3.06 & -2.72 & -3.05 & -2.73 & -3.04 & -2.85 & -3.23 & -3.17 \\
$\rm(H_2O)_2$    & -4.64 & -5.03 & -4.78 & -5.05 & -4.73 & -4.99 & -4.73 & -4.95 & -5.02 \\
Formic acid dimer & -17.52 & -18.76 & -17.92 & -18.80 & -17.71 & -18.53 & -17.61 & -18.31 & -18.80 \\
Formamide dimer & -14.34 & -15.70 & -14.72 & -15.68 & -14.61 & -15.50 & -14.62 & -15.35 & -16.12 \\
Uracil dimer & -18.46 & -20.23 & -18.96 & -20.21 & -18.84 & -20.01 & -18.66 & -19.58 & -20.69 \\
2-Pyridoxine -\\ 
2-Aminopyridine & -14.84 & -16.89 & -15.30 & -16.73 & -15.34 & -16.67 & -15.38 & -16.48 & -17.00 \\
Adenine-thymine WC & -13.95 & -16.16 & -14.45 & -16.02 & -14.49 & -15.94 & -14.58 & -15.75 & -16.74 \\
\multicolumn{10}{c}{Dispersion dominated complexes}\\
$\rm(CH_4)_2$      & 0.08 & -0.33 & -0.03 & -0.30 & -0.07 & -0.32 & -0.21 & -0.44 & -0.53 \\
$\rm(C_2H_4)_2$   & -0.22 & -1.35 & -0.49 & -1.30 & -0.57 & -1.32 & -0.82 & -1.41 & -1.50 \\
Benzene-$\rm CH_4$ & -0.09 & -1.28 & -0.39 & -1.20 & -0.52 & -1.28 & -0.77 & -1.38 & -1.45 \\
PD benzene dimer & 1.05 & -1.99 & 0.20 & -1.83 & -0.30 & -2.20 & -0.81 & -2.39 & -2.62 \\
Pyrazine dimer & -0.39 & -3.56 & -1.31 & -3.46 & -1.83 & -3.83 & -2.28 & -3.99 & -4.20 \\
Uracil dimer stack & -4.29 & -8.91 & -5.40 & -8.54 & -5.87 & -8.79 & -6.32 & -8.81 & -9.74 \\
Idole-benzene stack & 0.66 & -3.68 & -0.57 & -3.50 & -1.31 & -4.04 & -1.93 & -4.28 & -4.59 \\
Adenine-thymine stack & -4.03 & -10.40 & -5.74 & -10.09 & -6.57 & -10.63 & -7.24 & -10.88 & -11.66 \\
\multicolumn{10}{c}{Mixed complexes}\\
Ethene-ethine & -1.00 & -1.55 & -1.14 & -1.51 & -1.17 & -1.51 & -1.29 & -1.56 & -1.51 \\
Benzene-$\rm H_2O$ & -1.92 & -3.09 & -2.22 & -3.02 & -2.31 & -3.05 & -2.51 & -3.16 & -3.29 \\
Benzene-$\rm NH_3$ & -0.91 & -2.12 & -1.22 & -2.03 & -1.33 & -2.09 & -1.56 & -2.19 & -2.32 \\
Benzene-$\rm HCN$ & -3.11 & -4.61 & -3.55 & -4.59 & -3.66 & -4.63 & -3.90 & -4.76 & -4.55 \\
Benzene dimer T & -0.56 & -2.50 & -1.06 & -2.39 & -1.30 & -2.54 & -1.61 & -2.62 & -2.71 \\
Indole-benzene T & -2.75 & -5.31 & -3.47 & -5.21 & -3.80 & -5.42 & -4.18 & -5.64 & -5.62 \\
Phenol dimer & -4.54 & -6.79 & -5.12 & -6.69 & -5.31 & -6.77 & -5.48 & -6.61 & -7.09 \\
\hline
RMS  & 2.98 & 0.48 & 2.31 & 0.60 & 2.03 & 0.46 & 1.77 & 0.49 &  \\
MAD  & 2.39 & 0.36 & 1.84 & 0.45 & 1.66 & 0.36 & 1.44 & 0.36 &  \\
%standard dev & 10.01 & 0.07 & 6.07 & 0.18 & 3.19 & 0.10 & 1.94 & 0.16 & 1.37 & 0.08 & 1.08 & 0.12 &  \\
MD   & 7.63 & 1.26 & 5.92 & 1.57 & 5.09 & 1.03 & 4.42 & 1.11 &  \\
\hline
\hline
\end{tabular}\end{table}

\begin{table}\footnotesize
\caption{\label{s22acct}Dual basis resolution-of-the-identity double hybrid
aug-cc-pVTZ conformer energies for the
S22 test set of non-covalent bonded molecules in kcal/mol.  {\it Ab
initio} reference energies are the CBS($\Delta$a(DT)Z) values 
from Takatani {\it et al.}\cite{takatani2010}
}
\begin{tabular}{lrrrrrrrrr}
\hline
\hline
\multicolumn{1}{c}{aug-cc-pVTZ}           
& \multicolumn{2}{c}{B2-PLYP} & \multicolumn{2}{c}{B2GP-PLYP} & \multicolumn{2}{c}{DSD-BLYP} & \multicolumn{2}{c}{DSD-PBEP86}
& \multicolumn{1}{c}{\it ab initio}\\
%& \multicolumn{1}{c}{CBS($\Delta$a(DT)Z)}\\
& & \multicolumn{1}{c}{$-D3$} & & \multicolumn{1}{c}{$-D3$} & & \multicolumn{1}{c}{$-D3$} & & \multicolumn{1}{c}{$-D2$} \\
\hline
\multicolumn{10}{c}{H-bonded complexes}\\
$\rm (NH_3)_2$      & -2.58 & -3.06 & -2.75 & -3.08 & -2.79 & -3.09 & -2.90 & -3.29 & -3.17 \\
$\rm (H_2O)_2$      & -4.68 & -5.07 & -4.82 & -5.09 & -4.80 & -5.05 & -4.81 & -5.02 & -5.02 \\
Formic acid dimer   & -17.67 & -18.90 & -18.14 & -19.03 & -17.99 & -18.81 & -17.90 & -18.59 & -18.80 \\
Formamide dimer     & -14.60 & -15.97 & -15.06 & -16.02 & -15.01 & -15.91 & -15.01 & -15.74 & -16.12 \\
Uracil dimer        & -18.82 & -20.59 & -19.38 & -20.63 & -19.33 & -20.49 & -19.13 & -20.05 & -20.69 \\
2-pyridoxine - \\ 
2-aminopyridine     & -15.07 & -17.11 & -15.60 & -17.03 & -15.70 & -17.03 & -15.74 & -16.84 & -17.00 \\
Adenine thymine WC  & -14.27 & -16.48 & -14.85 & -16.42 & -14.95 & -16.40 & -15.04 & -16.21 & -16.74 \\
\multicolumn{10}{c}{Dispersion dominated complexes}\\
$\rm (CH_4)_2$      & 0.03 & -0.38 & -0.09 & -0.36 & -0.14 & -0.40 & -0.28 & -0.50 & -0.53 \\
$\rm (C_2H_4)_2$    & -0.32 & -1.44 & -0.61 & -1.42 & -0.72 & -1.47 & -0.97 & -1.56 & -1.50 \\
Benzene-$\rm CH_4$& -0.17 & -1.36 & -0.50 & -1.31 & -0.65 & -1.41 & -0.90 & -1.50 & -1.45 \\
PD Benzene dimer    & 0.71 & -2.33 & -0.21 & -2.25 & -0.78 & -2.68 & -1.28 & -2.86 & -2.62 \\
Pyrazine dimer      & -0.79 & -3.96 & -1.80 & -3.95 & -2.40 & -4.40 & -2.84 & -4.55 & -4.20 \\
Uracil dimer stack  & -4.84 & -9.47 & -6.09 & -9.23 & -6.66 & -9.59 & -7.11 & -9.59 & -9.74 \\
Indole-benzene stack& 0.19 & -4.15 & -1.15 & -4.08 & -1.97 & -4.70 & -2.58 & -4.94 & -4.59 \\
Adenine-thymine stack& -4.75 & -11.12 & -6.62 & -10.97 & -7.59 & -11.64 & -8.24 & -11.89 & -11.66 \\
\multicolumn{10}{c}{Mixed complexes}\\
Ethene-ethine       & -1.02 & -1.57 & -1.18 & -1.54 & -1.22 & -1.57 & -1.35 & -1.62 & -1.51 \\
Benzene-$\rm H_2O$  & -2.07 & -3.23 & -2.41 & -3.20 & -2.53 & -3.27 & -2.77 & -3.42 & -3.29 \\
Benzene-$\rm NH_3$  & -1.04 & -2.25 & -1.38 & -2.19 & -1.53 & -2.28 & -1.78 & -2.41 & -2.32 \\
Benzene-$\rm HCN$   & -3.14 & -4.63 & -3.61 & -4.66 & -3.76 & -4.73 & -4.02 & -4.88 & -4.55 \\
Benzene dimer T     & -0.68 & -2.62 & -1.23 & -2.56 & -1.51 & -2.75 & -1.82 & -2.82 & -2.71 \\
Indole-benzene T    & -2.87 & -5.42 & -3.65 & -5.39 & -4.03 & -5.66 & -4.42 & -5.88 & -5.62 \\
Phenol dimer        & -4.73 & -6.98 & -5.37 & -6.94 & -5.61 & -7.08 & -5.78 & -6.91 & -7.09 \\
\hline
RMS  & 2.70 & 0.21 & 1.96 & 0.27 & 1.62 & 0.13 & 1.36 & 0.27 &  \\
MAD  & 2.17 & 0.16 & 1.56 & 0.20 & 1.33 & 0.09 & 1.10 & 0.21 &  \\
%standard dev & 9.73 & 0.05 & 5.87 & 0.15 & 2.59 & 0.02 & 1.40 & 0.03 & 0.86 & 0.01 & 0.64 & 0.03 &  \\
MD   & 6.91 & 0.54 & 5.04 & 0.69 & 4.07 & 0.34 & 3.42 & 0.64 &  \\
\hline
\hline
\end{tabular}\end{table}

%\begin{figure}
%\includegraphics[width=\columnwidth]{timing-2.eps}
%\caption{\label{timing1}Relative timing speedup for the S22 test set between
%standard and dual basis DFT (black open bars) as well as standard and
%resolution-of-the-identity MP2 (red shaded bars).}
%\end{figure}

\end{document}